\documentclass[a4paper,10pt]{article}
\usepackage{a4wide}                 
\usepackage{amsmath}
\usepackage{amsthm}
\usepackage{amsfonts}
\usepackage{amssymb}
\usepackage{array}
\usepackage{graphicx}
\usepackage{listings}
\usepackage{multirow}               
\usepackage[longnamesfirst]{natbib}
\usepackage{bbm}

\theoremstyle{plain}

\theoremstyle{definition}
\newtheorem{example}{Example}

\theoremstyle{remark}

\renewcommand{\theenumi}{(\roman{enumi})}  

\DeclareMathOperator{\sgn}{sgn}
\DeclareMathOperator{\re}{Re}
\DeclareMathOperator{\im}{Im}
\DeclareMathOperator{\var}{Var}

\DeclareMathOperator{\supp}{supp}

\newcommand{\di}{\,\mathrm{d}}      
\newcommand{\eps}{\varepsilon}
\newcommand{\R}{\mathbb R}
\newcommand{\F}{\mathcal F}

\newcommand{\FA}{(FA) }
\newcommand{\SD}{(SD) }

\renewcommand{\tilde}{\widetilde}
\renewcommand{\hat}{\widehat}
\renewcommand{\theta}{\vartheta}

\title{Option calibration of exponential L\'evy models:\\ Confidence intervals and empirical results}
\author{Jakob S\"ohl\thanks{We thank Denis Belomestny and Markus Rei{\ss} for helpful comments and discussions. We are grateful for the comments and stimulating questions by two anonymous referees which have led to considerable improvements. This research was supported by the Deutsche Forschungsgemeinschaft through the SFB 649 ``Economic Risk''.}~\thanks{Humboldt--Universit\"at zu Berlin, Unter den Linden 6, 10099 Berlin, Germany, Tel.: +49 30 2093 3988, E--mail: soehl@math.hu--berlin.de} \and Mathias Trabs\footnotemark[1]~\thanks{Humboldt--Universit\"at zu Berlin, Unter den Linden 6, 10099 Berlin, Germany, Tel.: +49 30 2093 3988, E--mail: trabs@math.hu-berlin.de}}
\date{\today}
\begin{document}

\maketitle

\begin{abstract}
  Observing prices of European put and call options, we calibrate exponential L\'evy models nonparametrically. We discuss the efficient implementation of the spectral estimation procedures for L\'evy models of finite jump activity as well as for self--decomposable L\'evy models. Based on finite sample variances, confidence intervals are constructed for the volatility, for the drift and, pointwise, for the jump density. As demonstrated by simulations, these intervals perform well in terms of size and coverage probabilities. We compare the performance of the procedures for finite and infinite jump activity based on options on the German DAX index and find that both methods achieve good calibration results. The stability of the finite activity model is studied when the option prices are observed in a sequence of trading days.
\end{abstract}

\par
\noindent
\textbf{Keywords:} European option $\cdot$ Jump diffusion $\cdot$ Self--decomposability $\cdot$ Confidence sets $\cdot$ Nonlinear inverse problem $\cdot$ Spectral cut--off\\
\\
\textbf{MSC (2010):} 60G51 $\cdot$ 62G15 $\cdot$ 91B25\\
\\
\textbf{JEL Classification:} C14 $\cdot$ G13

\section{Introduction}
In recent years exponential L\'evy models are frequently used for the purpose of pricing and hedging. Assuming a constant and known riskless interest rate $r\geq0$ and an initial value $S_0>0$, these models describe the price of a stock by
\begin{equation}\label{eqModel}
  S_t=S_0e^{rt+X_t},\quad t\ge0,
\end{equation}
where $(X_t)_{t\ge0}$ is a L\'{e}vy process with characteristic triplet $(\sigma^2,\gamma,\nu)$. Thus jumps of the price process are taken into account and heavy tails in the returns are modeled appropriately. It has been shown that exponential L\'evy models are capable of reproducing not only the volatility smile but also the fact that it becomes more pronounced for shorter maturities. Hence, they are more adequate for recovering the stylized facts of financial time series than the classical model by \cite{blackScholes:1973}. To apply model~\eqref{eqModel}, for example, for derivative pricing, one has to infer the L\'evy triplet $(\sigma^2,\gamma,\nu)$ under the risk--neutral measure from observable data, since the triplet determines completely the distributional properties of the stock $S$. The estimation of the characteristics based on a finite sample of vanilla option prices is the aim of the present paper. In general an accurate calibration is corrupted by two error types, see \cite{cont}. First, the possible model misspecification is the deviation from the model, which we reduce by considering nonparametric models. Second, the calibration error is the deviation within the model, that we assess by means of confidence intervals.

Exponential L\'evy models are studied in a wide range of pricing problems, for instance by \citet{asmussen:2004,contVoltchkova:2005,ivanov:2007}. The calibration has mainly focused on parametric models, cf. \citet{barndorfnielsen:1998, eberlein1998,Carr:2002} and the references therein. First nonparametric calibration procedures for finite activity L\'evy models were proposed by \cite{contTankov:2004} as well as by \cite{reiss1:2006}. In these approaches no parametrization is assumed and thus the model misspecification is reduced. The method of \cite{trabs:2011} extends the spectral calibration to the infinite activity case, more precisely to self--decomposable L\'evy processes. Nonparametric confidence intervals and bands for L\'evy densities have been constructed by \cite{figueroa} based on high frequency observations. \cite{soehl:2012} derived asymptotic confidence sets for the calibration of the risk neutral measure and based on observations of option prices and not on historical data.\par

The calibration of a completely general L\'evy process might be too much to hope for. Therefore, we consider two submodels. Under the first setup, denoted by (FA), the process $X$ is assumed to be a jump--diffusion whose L\'evy measure $\nu$ has finite total mass. In the second case, which we refer to as (SD), we consider a self--decomposable L\'evy process without diffusion component, that is $\sigma=0$. In particular, in the second setting $\nu$ has infinite total mass and thus the two setups are non--overlapping. In both cases we do not assume that the L\'evy density $\nu$ belongs to some parametric, that is finite dimensional, class. Our estimators for $(\sigma^2,\gamma,\nu)$ in the two models \FA and \SD are constructed essentially as in \citet{reiss1:2006} and \citet{trabs:2011}, respectively, but some modifications are introduced which improve their numerical performance. As shown in simulations these improvements reduce the mean squared error of the estimators significantly. In contrast to the method by \cite{contTankov:2004} the spectral calibration is a straightforward algorithm, where no minimization problem has to be solved. Therefore, the methods are quite fast owing to the Fast Fourier transform (FFT). Whereas the above mentioned works focus on the asymptotic theory, we concentrate on the application of the method to realistic sample sizes. In a related framework of a jump--diffusion Libor model, \cite{belomestnySchoenmakers:2011} study the application of the spectral calibration method to finite sample data sets.\par

The construction of confidence intervals is based on the analysis of \cite{soehl:2012}, who derives asymptotic confidence sets in the finite activity case (FA). 
However, simulations with sample sizes as in available data show that these asymptotic confidence sets are too conservative.
To describe the behavior of the estimators more precisely, our confidence intervals use finite sample variances. Furthermore, this approach is extended to the self--decomposable scenario (SD). These intervals perform well in terms of size and coverage probabilities as demonstrated by simulations from the model by \cite{merton:1976} and from the variance gamma model, introduced by \citet{Madan:1990} and \citet{Madan:1998}.\par

We use data of vanilla options on the German DAX index to compare the finite activity model to the self--decomposable one. Considering options with different maturities, both models achieve good calibration results in the sense that the residuals between the given data and the calibrated model are small. Since the Blumenthal--Getoor index equals zero in our models, the calibration based on option data behaves quite differently from the case of high--frequency observations under the historical measure, where \cite{aitSahaliaJacod:2009} find evidence that the Blumenthal--Getoor index is larger than one. Applying the calibration to a sequence of trading days, we obtain the evolution of the model parameters in time. The estimators seem to be stable with respect to the spot time.\par

This paper is organized as follows: In Section~\ref{sec:model} we state precisely the models \FA and \SD and describe the general estimation method. The explicit estimators for the finite activity case and the self--decomposable case are constructed in Sections~\ref{sec:finiteActiv} and \ref{sec:sd}, respectively. In Section~\ref{sec:confidence} the confidence intervals are derived and their performance is assessed in simulations. We apply the methods to data and discuss our results in Section~\ref{sec:empirical}. We conclude in Section~\ref{sec:conclusion}. The more technical part of determining the finite sample variances is deferred to the appendix.

\section{Model and estimation principle}\label{sec:model}
Let us first recall some basic properties of the L\'evy process $(X_t)_{t\ge0}$. By definition it is a stochastically continuous processes which starts at zero and which has stationary and independent increments. Due to the L\'evy--It\^{o} decomposition \citep[Thm. 19.3]{sato:1999}, $X_t$ can be written as the sum of a Brownian motion with drift $\sigma W_t+\gamma t$ and an independent pure jump process $J_t$, where $\sigma\ge0$ denotes the volatility and $\gamma\in\R$ the drift. The jump part can be completely described by the jump measure $\nu$ on the real line. Throughout, we assume
\newcounter{ass}
\begin{enumerate}
  \renewcommand{\theenumi}{(A\arabic{ass})}
  \stepcounter{ass}\item\label{ass1} $\nu$ is absolutely continuous. Abusing notation, we denote its Lebesgue density likewise by $\nu:\R\to\R_+$.
  \stepcounter{ass}\item\label{ass2} $\int_\R(|x|\wedge1)\nu(x)\di x<\infty$.
\end{enumerate}
Owing to \ref{ass2}, the jump component $J_t$ has finite variation. Therefore, the characteristic function of $X_t$ is given by the L\'{e}vy--Khintchine representation \citep[Thm. 8.1]{sato:1999}
\begin{equation}\label{eqLevyKhint}
  \varphi_T(u):=\mathbb E[e^{iuX_T}]=e^{T\psi(u)}\quad\text{where }\quad \psi(u):=-\frac{\sigma^2}{2}u^2+i\gamma u+\int_\R(e^{iux}-1)\nu(x)\di x.
\end{equation}
The L\'evy process is uniquely determined by the so called characteristic triplet $(\sigma^2,\gamma,\nu)$ and thus calibrating the exponential L\'evy model~\eqref{eqModel} reduces to estimating the two one--dimensional parameters $\sigma^2$ and $\gamma$ as well as the density $\nu$ from an infinite dimensional parameter space. However, the characteristic triplet depends on the underlying measure. Since we are interested in pricing and hedging purposes, we consider throughout the risk neutral measure under which the discounted process $e^{X_t}$ is a martingale. Therefore, $\mathbb E[e^{X_t}]=1, t\geq0,$ which is equivalent to the martingale condition
\begin{equation}\label{eqMartCond}
  \frac{\sigma^2}{2}+\gamma+\int_{-\infty}^\infty(e^x-1)\nu(\di x)=0.
\end{equation}
So far, nonparametric calibration methods exist in two different setups:
\begin{enumerate}
  \item[\FA] Assumptions ~\ref{ass1} holds and Assumption \ref{ass2} is replaced by the stronger assumption of finite activity $\lambda:=\int_{\R}\nu(x)\di x<\infty$ \citep{contTankov:2004,reiss1:2006,soehl:2012}.
  \item[\SD] $X_t$ is self--decomposable with $\sigma=0$ that is $\nu$ can be characterized by $\nu(\di x)=k(x)/|x|\di x$ for $x\in\R\setminus\{0\}$, where $k$ is increasing on $\R_-$ and decreasing on $\R_+$. Additionally, $\alpha:=k(0+)+k(0-)$ is assumed to be finite \citep{trabs:2011}.
\end{enumerate}
Note that in the \SD setting Assumptions~\ref{ass1} and \ref{ass2} are automatically satisfied. The function~$k$ with the above monotonicity properties is called k-function. \citet{trabs:2011} considers a more general class of L\'evy processes where $k$ does not need to fulfill these monotonicity properties. However, we will see that the class of self--decomposable processes is already rich enough to calibrate the model \eqref{eqModel} well.

Typical parametric submodels of \FA and \SD are given by Examples~\ref{exMerton} and \ref{exVG}, respectively. We will use them to study the performance of estimation methods in simulations.
\begin{example}[Merton model]\label{exMerton}
  \cite{merton:1976} introduced the first exponential L\'evy model. Therein, the jumps are normally distributed with intensity $\lambda>0$:
  \[
    \nu(x)=\frac{\lambda}{\sqrt{2\pi}v}\exp\left(-\frac{(x-\eta)^2}{2v^2}\right),\quad x\in\mathbb{R}.
  \]
  A realistic choice of the parameters is $\eta=-0.1$, $v=0.2$ and $\lambda=5$. Together with the volatility $\sigma=0.1$ this determines the drift to be $\gamma=0.379$ using the martingale condition \eqref{eqMartCond}.
\end{example}

\begin{example}[Variance gamma model]\label{exVG}
  Let $(W_t)$ be a standard Brownian motion and $(G_t)$ an independent Gamma process with mean rate one and variance rate $\rho$ that is $G_t\sim\Gamma(t/\rho,1/\rho)$. \cite{Madan:1990} defined the variance gamma process with parameters $\sigma, \rho$ and $\theta$ as the time changed Brownian motion with drift $X_t=\theta G_t+\sigma W_{G_t}, t\geq0$. 
  This is a model with infinite jump activity and Blumenthal--Getoor index zero.
  The characteristic function and the k--function of $(X_t)$ are given by
  \begin{align*}
    \varphi_t(u)&=(1+i\theta\rho u+\sigma^2\rho u^2/2)^{-t/\rho}\quad\text{and}\\
    k_{VG}(x)&=\frac{1}{\rho}e^{x/\eta_m}\mathbf1_{\{x<0\}}(x)+\frac{1}{\rho}e^{-x/\eta_p}\mathbf1_{\{x\geq0\}}(x),\quad u,x\in\mathbb R,
 \end{align*}
  with $\eta_{p}:=\sqrt{\theta^2\rho^2/4+\sigma^2\rho/2}+\theta\rho/2$ and $\eta_{m}:=\sqrt{\theta^2\rho^2/4+\sigma^2\rho/2}-\theta\rho/2$, respectively. In our simulations we use the parameters $\sigma=1.2, \rho=0.2$ and $\theta=-0.15$. The value of $\gamma=0.141$ is given by the martingale condition again. These choices imply $\alpha=k_{VG}(0+)+k_{VG}(0-)=10$.
\end{example}
Since we want to estimate the model parameters under the risk neutral measure, the procedure is based on observing prices of vanilla options. Throughout, we measure the time in years. Let us fix a maturity $T>0$, define the negative log--moneyness $x:=\log(K/S_0)-rT$ and denote call and put prices by $\mathcal{C}(x,T)=S_0\mathbb E[(e^{X_T}-e^x)_+]$ and $\mathcal{P}(x,T)=S_0\mathbb E[(e^x-e^{X_T})_+]$, respectively. In terms of the option function
\[
  \mathcal{O}(x):=\begin{cases}
                     \displaystyle S_0^{-1}\mathcal{C}(x,T), & \quad x\ge0,\\
                     \displaystyle S_0^{-1}\mathcal{P}(x,T), & \quad x<0,
                  \end{cases}
\]
our observations are given by
\begin{equation}\label{eqObservation}
  O_j=\mathcal{O}(x_j)+\delta_j \eps_j, \quad j=1,\dots,N,
\end{equation}
with noise levels $\delta_j>0$ and independent, centered errors $\eps_j$, satisfying $\var(\eps_j)=1$ as well as $\sup_j\mathbb E[\eps_j^4]<\infty$. The observation errors are due to the bid--ask spread and other market frictions. For simplicity, we assume $(\delta_j)_{j=1,\dots,N}$ to be known. Otherwise, the noise levels can be estimated on an independent data set, for instance, from market data which contain separately bid and ask prices. Note that since the L\'evy density $\nu$ is an infinite--dimensional object, the triplet $(\sigma^2,\gamma,\nu)$ cannot be inferred from the market price of just one vanilla option as the volatility parameter in the Black--Scholes model. The more prices $O_j$ are observed for different strikes $x_j$, the more accurate the estimation will be. To construct the estimators of the L\'evy triplet, we apply the L\'{e}vy--Khintchine representation \eqref{eqLevyKhint} and the pricing formula by \citet{carrMadan:1999}
\begin{equation}\label{pricingFormula}
  \mathcal{FO}(u):=\int_{-\infty}^{\infty}e^{iux}\mathcal{O}(x)\di x=\frac{1-\varphi_T(u-i)}{u(u-i)}.
\end{equation}
Note that the latter equation extends to all complex numbers in the strip $\{u\in\mathbb C|\im (u)\in[0,1]\}$ since there the characteristic function $\varphi_T(u-i)$ is finite by the exponential moment of $X_T$, which is implied by the martingale condition \eqref{eqMartCond}. We obtain
\begin{align}
  \psi(u)=&\frac{1}{T}\log(1-u(u+i)\mathcal{FO}(u+i)),\label{eqPsi}\\
  \psi_{-i}(u):=\psi(u-i)=&\frac{1}{T}\log(1+iu(1+iu)\mathcal{FO}(u))\label{eqPsi-i}.
\end{align}
Through curve fitting to $(x_j,O_j)_{j=1,\dots,N}$, we obtain an empirical versions $\tilde O$ of the option function and subsequently, through a plug--in approach, empirical versions $\tilde\psi$ and $\tilde\psi_{-i}$ of the characteristic exponents. While the theoretical results \citep{reiss1:2006, trabs:2011} concentrate on a linear interpolation of the observation, an additional smoothing by using B--splines of degree two might improve the estimators. In Section~\ref{sec:FAsim} we provide simulations with both interpolation methods to investigate the practical influence.\par

Given $\tilde\psi$, we can estimate the characteristics of the process from the spectral representation. The procedures of \citet{reiss1:2006} as well as \citet{trabs:2011} rely on the identity \eqref{eqPsi-i} which looks more convenient because it uses directly the option function. The identity \eqref{eqPsi} uses an exponentially scaled option function since $\mathcal{FO}(u+i)=\mathcal{F}[e^{-x}\mathcal O(x)](u)$. However, in \eqref{eqPsi-i} the characteristic exponent
is shifted by $-i$, which leads to estimators of exponentially scaled versions of the jump density $\nu$ and of the k-function~$k$, respectively. Therefore, we will use it only to estimate the one--dimensional parameters of the models. According to the idea of \cite{reiss2:2006}, equation \eqref{eqPsi} allows to estimate immediately the nonparametric objects $\nu$ and $k$. Regularization of the procedure is achieved by cutting off frequencies larger than a regularization parameter $U>0$. Since \FA and \SD need to be considered separately, the precise estimators are given in Sections~\ref{sec:finiteActiv} and \ref{sec:sd}. Note that in both cases correction steps are necessary to satisfy non--negativity of the jump density and the martingale condition \eqref{eqMartCond} \citep[see][for details]{soehlTrabs:2012b}. If the latter one would be violated, the right--hand side of the pricing formula \eqref{pricingFormula} could have a singularity at zero and thus we could not apply the inverse Fourier transform to obtain an option function from the calibration.\par

A critical question is the choice of the regularization parameter $U$. As a benchmark, we use in simulations an oracle cut--off value, that is $U$ minimizes the discrepancy between the estimators and the true values of $\sigma^2,\gamma$ and $\nu$ measured in an $L^2$--loss. To calibrate real data, we employ the simple least squares approach
\begin{equation}\label{eqRSS}
      U^*:=\operatorname{arginf}_U RSS(U)\quad\text{ with the residual sum of squares } RSS(U):=\sum\limits_{j=1}^N |\hat{\mathcal O}_U(x_j)-O_j|^2,
\end{equation}
where $\hat{\mathcal O}_U$ is the option function corresponding to the L\'evy triplet estimated by means of the cut--off value~$U$. We determine $\hat{\mathcal O}_U$ by the pricing formula \eqref{pricingFormula} and L\'evy-Khintchine representation \eqref{eqLevyKhint}, in which we plug in the estimators obtained by using the cut--off value~$U$. 
The estimated option function $\hat{\mathcal O}_U$ can be computed efficiently for each $U$ so that the numerical effort of finding $U^*$ is mainly determined by the minimization algorithm used to solve~\eqref{eqRSS}.
From theoretical consideration a penalty term, as used by \cite{reiss2:2006}, is necessary to avoid an over--fitting, that is not to choose $U$ too large. Nevertheless, our practical experience with this method shows that the above mentioned correction steps, which are not included in the theory, lead to an auto--penalization: Using large cut--off values, the stochastic error in the estimators becomes large. This leads to high fluctuations of the nonparametric part and thus the correction has an increasing effect. Hence, the difference between $\tilde{\mathcal O}$ and $\hat{\mathcal O}_U$ becomes larger if $U$ is too high and thus the residual sum of squares increases, too. In particular, imposing the jump density to be nonnegative implies a shape constraint on the state price density which is basically the second derivative of the option function. Therefore, the least squares choice of the tuning parameter works well at least for small noise levels.

The approach to minimize the calibration error was also applied by \cite{belomestnySchoenmakers:2011}. Alternative data--driven choices of the cut--off value $U$ are the ``quasi-optimality'' approach which was studied by \citet{bauerReiss:2008} and which was applied by \citet{belomestny:2011} or the use of a preestimator as proposed by \citet{trabs:2011}. However, we will consider only the least squares approach which performs well in our application.

\section{The finite activity case}\label{sec:finiteActiv}
\subsection{The estimators}
In the \FA setup we deduce from \eqref{eqLevyKhint} and \eqref{eqPsi-i} the identity
\[\psi_{-i}(u)=-\frac{\sigma^2}{2}u^2+i(\sigma^2+\gamma)u +(\sigma^2/2+\gamma-\lambda)+\mathcal{F}\mu(u)\quad\text{with } \mu(x):=e^x\nu(x).\]
The estimators of the parameters are defined by \citet{reiss1:2006} as follows:
\begin{align}
\hat{\sigma}^2 &:= \int_{-U}^{U}\re(\tilde\psi_{-i}(u))w_{\sigma}^{U}(u)\mathrm{d}u,\label{sigmaHat}\\
\hat\gamma_{fa} &:= -\hat{\sigma}^2 +\int_{-U}^{U}\im(\tilde\psi_{-i}(u))w_{\gamma_{fa}}^{U}(u)\mathrm{d}u,\label{gammaHat}\\
\hat{\lambda} &:= \frac{\hat{\sigma}^2}{2}+\hat\gamma_{fa} -\int_{-U}^{U}\re(\tilde\psi_{-i}(u))w_{\lambda}^{U}(u)\mathrm{d}u\label{lambdaHat},
\end{align}
where $w_\sigma^{U}$, $w_{\gamma_{fa}}^{U}$ and $w_\lambda^{U}$ are suitable weight functions and $\tilde\psi_{-i}$ is the empirical version of $\psi_{-i}$. To avoid ambiguity, the estimator of $\gamma$ has an additional subscript denoting the model in which the estimator is defined.
The estimators in \eqref{sigmaHat},\eqref{gammaHat} and \eqref{lambdaHat} can be understood as weighted $L^2$--projections of $\tilde\psi_{-i}$ onto the space of quadratic polynomials. In this sense the estimators arise naturally as a solution of a weighted least squares problem.
However, the optimal weight depends not only on the unknown heteroscedacity in the frequency domain but also on the unknown function $\mathcal{F}\mu$, so we do not pursue this approach here. Instead we construct the weight functions $w_\sigma^{U}$, $w_{\gamma_{fa}}^{U}$ and $w_\lambda^{U}$ directly as \cite{reiss2:2006} but propose different weight functions. The idea is that the noise is particularly high in the high frequencies and thus it is desirable to assign less weight to the high frequencies.
A smooth transition of the weight functions to zero at the cut--off value improves the numerical results significantly. Therefore, we would like the weight function and its first two derivatives to be zero at the cut--off value. With the side conditions on the weight functions this leads to the following polynomials:
\begin{align*}
  w_\sigma^U(u)&:=\frac{c_\sigma}{U^3}\Big((2s+1)\left(\frac{u}{U}\right)^{2s}- 4(2s+3)\left(\frac{u}{U}\right)^{2s+2}+ 6(2s+5)\left(\frac{u}{U}\right)^{2s+4}\\
      &\qquad -4(2s+7)\left(\frac{u}{U}\right)^{2s+6} +(2s+9)\left(\frac{u}{U}\right)^{2s+8}\Big)\mathbbm 1_{[-U,U]}(u),\\
  w_{\gamma_{fa}}^U(u)&:=\frac{c_{\gamma_{fa}}}{U^2}\left(\left(\frac{u}{U}\right)^{2s+1}- 3\left(\frac{u}{U}\right)^{2s+3}+ 3\left(\frac{u}{U}\right)^{2s+5} -\left(\frac{u}{U}\right)^{2s+7}\right)\mathbbm 1_{[-U,U]}(u),\\
  w_\lambda^U(u)&:=\frac{c_\lambda}{U}\Big((2s+3)\left(\frac{u}{U}\right)^{2s}- 4(2s+5)\left(\frac{u}{U}\right)^{2s+2}+ 6(2s+7)\left(\frac{u}{U}\right)^{2s+4}\\
      &\qquad -4(2s+9)\left(\frac{u}{U}\right)^{2s+6} +(2s+11)\left(\frac{u}{U}\right)^{2s+8}\Big)\mathbbm 1_{[-U,U]}(u),
\end{align*}
where all three functions equal zero outside $[-U,U]$. The constants $c_\sigma, c_{\gamma_{fa}},c_\lambda\in\R$ are determined by the normalization conditions
\[
  \int_{-U}^Uu^2w_\sigma^U(u)\di u=-2,\quad
  \int_{-U}^Uuw_{\gamma_{fa}}^U(u)\di u=1\quad\text{and}\quad
  \int_{-U}^Uw_{\lambda}^U(u)\di u=1.
\]
The parameter $s$ reflects the a priori knowledge about the smoothness of $\nu$ and can be chosen equal to two. The gain of the new weight functions is discussed in Section~\ref{sec:FAsim}.

To estimate directly the jump density $\nu$ and not only the exponential scaled version $\mu$, we use $\psi$ instead of $\psi_{-i}$ as discussed above. Therefore, we define the estimator
\begin{align}
  \hat{\nu}(x)&:=\mathcal{F}^{-1}\biggl[\left(\tilde{\psi}(u)+\frac{\hat{\sigma}^2}{2}u^2-i\hat\gamma_{fa} u+ \hat \lambda \right)w_{\nu}^{U}(u)\biggr](x),\label{nuHat}
\end{align}
where $\tilde\psi$ is the empirical version of $\psi$ and $w_\nu^U$ is a flat top kernel with support $[-U,U]$:
\begin{equation}\label{flatTopKernel}
  w_\nu^U(u):=F\left(\frac{u}{U}\right)\quad \text{with}\quad F(u):=\begin{cases}
    1, & |u|\leq0.05,\\
    \exp\left(\frac{-\exp(-(|u|-0.05)^{-2})}{(|u|-1)^2}\right), & 0.05<|u|<1,\\
    0, & |u|\geq1.
  \end{cases}
\end{equation}
To evaluate the integrals in \eqref{sigmaHat} to \eqref{lambdaHat}, it suffices to apply the trapezoidal rule. The inverse Fourier transformation in \eqref{nuHat} can be efficiently computed using the FFT--algorithm. Therefore, depending on the interpolation method which is applied to obtain $\tilde O$, the whole estimation procedure is very fast. Finally, we note that the cut--off value can be chosen differently for each quantity $\sigma^2,\gamma,\lambda$ and $\nu$. A documentation of the implementation in R can be found in \cite{soehlTrabs:2012b}.

\subsection{Simulations}\label{sec:FAsim}
Let us first describe the setting of all of our simulations. In view of the higher concentration of European options at the money, the design points $\{x_1,\dots,x_N\}$ are chosen to be the $k/(N+1)$--quantiles, $k=1,\dots,N,$ of a normal distribution with mean zero and variance $1/2$. The observations $O_j$ are computed from the characteristic function $\varphi_T$ using the fast Fourier transform. The additive noise consists of independent, normal and centered random variables with variance $|\tau\mathcal O(x_j)|^2$ for some relative noise level $\tau>0$. By choosing the sample size $N$ and the deviation parameter $\tau$, we determine the noise level of the observations. According to the existing theoretical results, it is well measured by the quantity
\[
    \eps:=\Delta^{3/2}+\Delta^{1/2}\|\delta\|_{l^\infty}\quad\text{with}\quad \Delta:=\max_{j=2,\dots,N}(x_j-x_{j-1}),
\]
which takes the interpolation error and the stochastic error into account. The interest rate and time to maturity are set to $r=0.06$ and $T=0.25$, respectively.\par
\begin{figure}[tp]\centering
  \includegraphics[width=9cm]{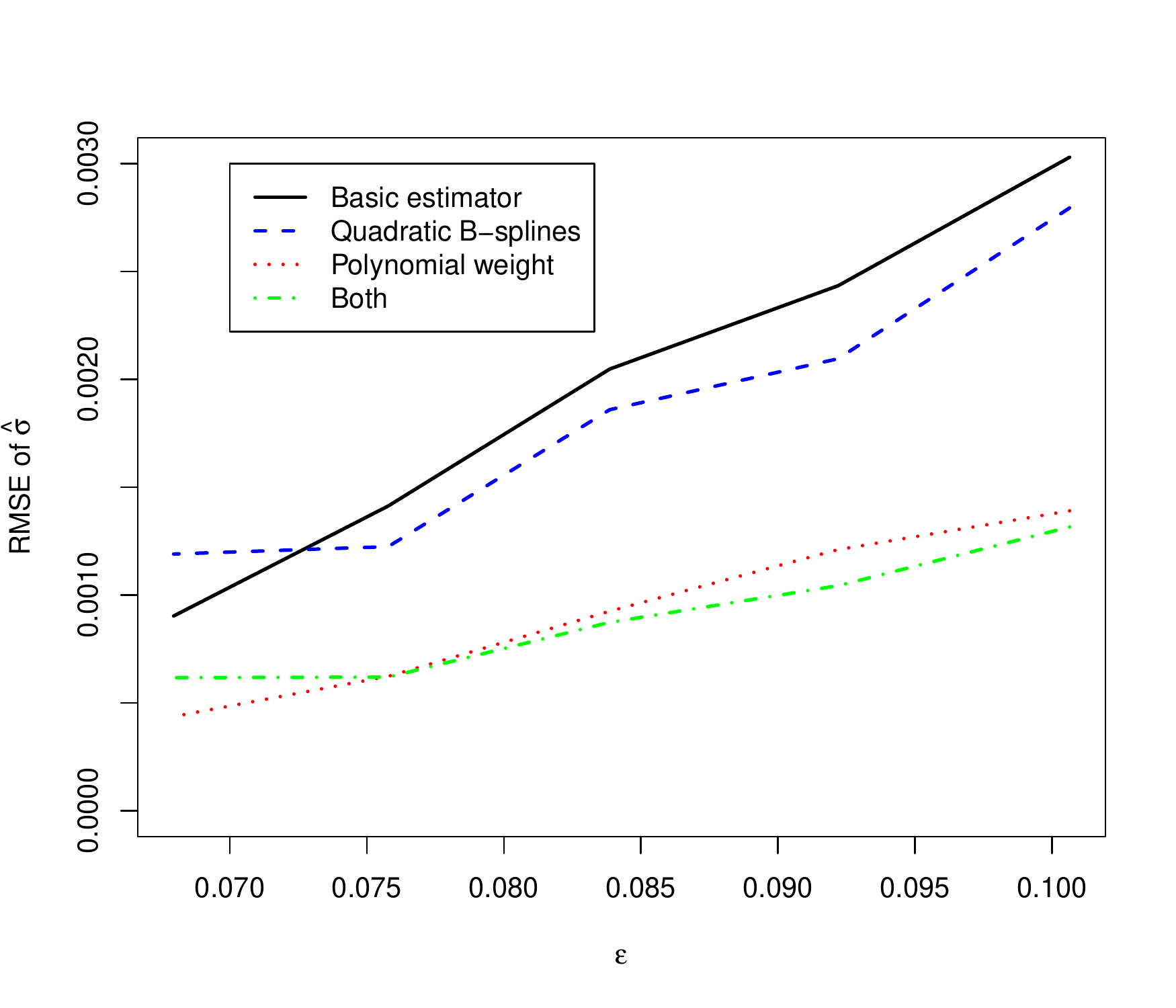}
  \caption{RMSE of $\hat\sigma$ for different noise levels with 500 Monte--Carlo iterations in each case. Usage of the linear and quadratic spline interpolation as well as usage of the weight functions by \cite{reiss1:2006} and the polynomial weight functions.}\label{figLinVsQuad}
\end{figure}

Using the Merton model with the parameters of Example~\ref{exMerton}, we investigate the practical influence of two aspects of the procedure, which are mentioned above. The interpolation of the data $(x_j,O_j)$ with linear B--splines is compared to the use of quadratic B--splines. The latter preprocessing is an additional smoothing of the data, which achieves significant gains for higher noise levels. The other point of interest is the choice of the weight functions. Since it is known from the theory that the noise affects mainly the high frequencies, the polynomial weight functions greatly reduce the variance of the estimator. These improvements are illustrated in Figure~\ref{figLinVsQuad}: In the case of $\hat\sigma$ we approximate the root mean squared error (RMSE) $\sqrt{\mathbb E[|\hat\sigma-\sigma|^2]}$ using 500 Monte--Carlo iterations with and without quadratic splines and polynomial weight functions, respectively. This is done for different noise levels, whereby $\tau$ decreases from 0.03 to 0.015 and $N$ increases from 50 to 400, simultaneously. Further simulation results, in particular for estimating the jump density, can be found in Section~\ref{sec:confidence}.

\section{The self--decomposable framework}\label{sec:sd}

Recall that $\sigma=0$ is assumed in the \SD setting. While the Blumenthal--Getoor index is zero in this case the parameter $\alpha$ describes the degree of activity of the process on a finer scale. To calibrate the self--decomposable model, we need a different representation of $\psi_{-i}$ than before because of the infinite activity of these processes. Applying Fubini's theorem to \eqref{eqLevyKhint}, we obtain $\psi_{-i}(u)=i\gamma u+\gamma+\int_0^1i(u-i)\mathcal F[\sgn(x)k(x)]((u-i)t)\di t, u\in\R,$ where the Fourier transform decays slowly since $\sgn(x)k(x)$ is not continuous at zero. \citet[Prop. 2.2]{trabs:2011} showed that decomposing $\sgn(x)k(x)$ into a nonsmooth and a smooth part yields for $u\neq0$
\begin{align}\label{eqPsiSD}
  \psi_{-i}(u)&=D(u)+i\gamma u-\alpha\log(|u|)+\sum\limits_{j=1}^{2s-2}\frac{i^j(j-1)!\alpha_j}{u^j}+\rho(u),
\end{align}
where $2s$ is the smoothness of $k$ away from zero, $\alpha_j:=k^{(j)}(0+)+k^{(j)}(0-)$ for $j=1,\dots,2s-2$, the function $D$ is constant on the real half lines and the remainder $\rho$ corresponds to the smooth part of $\sgn(x)k(x)$ and thus satisfies $\|u^{2s-1}\rho(u)\|_\infty<\infty$. Owing to the polynomial decay of $\rho$, estimators of $\gamma$ and $\alpha$ can be defined analogously to Section~\ref{sec:finiteActiv}, filtering the coefficient of the linear term and of the logarithmic term in \eqref{eqPsiSD}, respectively:
\begin{align*}
  \hat\gamma_{sd}&:=\int_{-U}^U\im(\tilde\psi_{-i}(u))w_{\gamma_{sd}}^U(u)\di u,\\
  \hat\alpha&:=\int_{-U}^U\re(\tilde\psi_{-i}(u))w_{\alpha}^U(u)\di u
\end{align*}
with polynomial weight functions
\begin{align*}
  w_{\gamma_{sd}}^U(u)=\frac{1}{U^2}\sum_{k=0}^{s+1}a_k\Big(\frac{u}{U}\Big)^{2(k+s)-1}\quad\text{and}\quad
  w_\alpha^U(u)=\frac{1}{U}\sum_{k=0}^{s+1}b_k\Big(\frac{u}{U}\Big)^{2(k+s)},
\end{align*}
where the coefficients $a_k,b_k\in\R$ are determined by
\begin{align*}
  \int_0^Uuw_{\gamma_{sd}}^U(u)\di u&=\frac{1}{2},&\quad
  \int_0^Uw_{\gamma_{sd}}^U(u)\di u&=0\quad\text{and}&\quad
  \int_0^Uu^{-2l+1}w_{\gamma_{sd}}^U(u)\di u&=0,\\
  \int_0^U\log(|u|)w_\alpha^U(u)\di u&=-\frac{1}{2},&\quad
  \int_0^Uw_\alpha^U(u)\di u&=0\quad\text{and}&\quad
  \int_0^Uu^{-2l}w_\alpha^U(u)\di u&=0,
\end{align*}
for $l=1,\dots,s-1$.
These integral conditions lead to a system of linear equations which can be solved analytically as well as numerically. To estimate the k-function, we combine the method of \cite{trabs:2011} with the approach by \cite{reiss2:2006}. From \eqref{eqLevyKhint}, Fubini's theorem and \eqref{eqPsi} follows
\begin{equation}\label{eqPsiPrim}
  \psi'(u)=i\gamma+i\mathcal F[\sgn\cdot k](u)
  =\frac{(u-iu^2)\F[x\mathcal O](u+i)-(2u+i)\F\mathcal O(u+i)}{T(1-u(u+i)\F\mathcal O(u+i))}.
\end{equation}
Let $\tilde\psi'$ be the empirical version of $\psi'$ obtained by substituting $\mathcal{O}$ by $\tilde O$ in \eqref{eqPsiPrim}. Since we know the position of the jump of $k$, the application of a one--side kernel function allows to estimate the k--function on the whole real line. We define
\begin{align*}
  \hat k(x):=\begin{cases}
                 \displaystyle\mathcal F^{-1}\big[(-\hat\gamma_{sd}-i\tilde\psi'(u))\mathcal FW_k(u/U)\big](x),\quad&x>0,\\
                 \displaystyle\mathcal F^{-1}\big[(\hat\gamma_{sd}+ i\tilde\psi'(u))\mathcal FW_k(-u/U)\big](x),\quad&x<0,
               \end{cases}
\end{align*}
with a one--sided kernel function
\[
  W_k(x):=\Big(\sum_{m=0}^{2s}c_mx^m\Big)F(x+1),\quad x\in\R,
\]
where $F$ is the flat top kernel defined in \eqref{flatTopKernel}, thus $\supp W_k=[-2,0]$, and the coefficients $c_k\in\R$ are chosen such that
\[
  \int W_k=1,\quad\int x^lW_k(x)\di x=0\quad\text{ for }l=1,\dots,2s-1.
\]
Again, the coefficients are given by a system of linear equations, which can be solved numerically. Because the kernel $W_k$ has to be one--sided, it cannot have compact support in the Fourier domain. Hence, to ensure that there are no large stochastic errors in $\tilde\psi'(u)$ for large $u\in\R$, a truncation in the spirit of \citet[p. 7]{trabs:2011} might be reasonable. To obtain an estimator in the class of self--decomposable processes, we have to ensure the necessary monotonicity of the k--function. Therefore, we apply
rearrangement, which is a general procedure to transform a function into a monotone function.
With some arbitrary large constant $C>0$, the rearranged estimator is then given by
\begin{equation}
  \hat k^*(x):=\begin{cases}
                 \displaystyle \inf\Big\{y\in\mathbb R_+\Big|\int_0^C\mathbf 1_{\{\hat k(z)\ge y\}}\di z\le x\Big\},\quad&x\in(0,C],\\
                 \displaystyle \inf\Big\{y\in\mathbb R_+\Big|\int_0^C\mathbf 1_{\{\hat k(-z)\ge y\}}\di z\le |x|\Big\},\quad&x\in[-C,0),\\
                 0,&\text{otherwise}.
               \end{cases}
\end{equation}
In the sequel, we identify $\hat k$ with its rearranged version $\hat k^*$, since we are interested only in the calibration using self-decomposable processes. For the application of this method to simulations and to real data, we refer to Sections~\ref{sec:confidence} and \ref{sec:empirical}.

 \section{Confidence intervals}\label{sec:confidence}

\citet{soehl:2012} shows asymptotic normality of the estimators in the \FA setup. These result may be used to construct confidence intervals in both models, the finite activity case and the self--decomposable one. Let us consider $\hat\sigma^2$ first. All other parameters can be treated similarly. As usual in nonparametric statistics the estimation error $\hat\sigma^2-\sigma^2$ decomposes into a deterministic approximation error and a stochastic part. The choice of the cut--off value $U$ allows a trade--off between these two errors. In order to construct confidence intervals, the cut--off value $U$ is chosen large enough such that the bias is asymptotically negligible. Due to this undersmoothing, we can approximate the estimation error by
\begin{equation}\label{error_sigma}
  \hat\sigma^2-\sigma^2
  \approx\int_{-U}^{U}\re(\Delta\tilde\psi_{-i}(u))w_\sigma^U(u)\di u
\end{equation}
with $\Delta\tilde\psi_{-i}:=\tilde\psi_{-i}-\psi_{-i}$.
The term $\Delta\tilde\psi_{-i}(u)$ is a logarithm, which we approximate by
\begin{align}\label{eqLinStochErr}
  \Delta\tilde\psi_{-i}(u)=\frac{1}{T}\log\Big(\frac{1+iu(1+iu)\F\tilde{\mathcal O}(u)}{1+iu(1+iu)\F\mathcal O(u)}\Big)
  \approx\frac{iu(1+iu)}{T\varphi_T(u-i)}(\F\tilde{\mathcal O}-\F\mathcal O)(u)
\end{align}
using $\log(1+x)\approx x$ for small $x$.
We apply the approximation \eqref{eqLinStochErr} to the right--hand side of \eqref{error_sigma} and call the resulting term linearized stochastic error. Confidence intervals may be constructed in two different ways. They can be derived either from the asymptotic variance or from the finite sample variance of the linearized stochastic errors. We will follow the second approach. Nevertheless, the confidence intervals are asymptotic in the sense that the approximation errors and the remainder terms of the stochastic errors are considered as negligible. \citet[Sects. 6.3 and 6.4]{soehl:2012} determines exact conditions under which both additional errors vanish asymptotically.\par

To analyze the deviation $\F\tilde{\mathcal O}-\F\mathcal O$ in the linearized stochastic error, we assume that the noise levels of the observations \eqref{eqObservation} are given by the values $\delta_j=\delta(x_j)$, $j=1,\dots,N$, of some function $\delta:\mathbb{R}\to\mathbb{R}_{+}$. The observation points are assumed to be the quantiles $x_j=H^{-1}(j/(N+1))$, $j=1,\dots,N$, of a distribution with c.d.f. $H:\mathbb{R}\to[0,1]$ and p.d.f. $h$. For the definition of the confidence intervals we need the generalized noise level
\begin{equation}\label{varrho}
\varrho(x)=\delta(x)/\sqrt{h(x)},
\end{equation}
which incorporates the noise of the observations as well as their distribution. Instead of assuming that the observation points are given by the quantiles of $h$ one may also assume that the observation points are sampled randomly from the density $h$. On these conditions \cite{brownLow:1996} showed the asymptotic equivalence in the sense of Le Cam of the nonparametric regression model~\eqref{eqObservation} and the Gaussian white noise model $\di Z(x)=\mathcal O(x)\di x+N^{-1/2}\varrho(x)\di W(x)$ with a two--sided Brownian motion $W$. More details on this equivalence can be found in the papers by \cite{soehl:2012} and \citet[Supplement]{trabs:2011}. $Z$ is an empirical version of the antiderivative of $\mathcal{O}$. In that sense we define $\F\tilde{\mathcal{O}}(u):=\F[\di Z](u)=\F\mathcal{O}(u)+N^{-1/2}\int_{\R}e^{iux}\varrho(x)\di W(x)$.
Combining \eqref{eqLinStochErr} with this asymptotic equivalence, we can approximate
\begin{align*}
  \Delta\tilde\psi_{-i}\approx\frac{1}{\sqrt N}\mathcal L(u):=\frac{1}{\sqrt N}\frac{iu(1+iu)}{T\varphi_T(u-i)}\int_\R e^{iux}\varrho(x)\di W(x).
\end{align*}
Defining $f_{\sigma,U}(u):=w_\sigma^U(u)iu(1+iu)/(T\varphi_T(u-i))$, the above considerations and exchanging the order of the integrals yield
\[
  \hat\sigma^2-\sigma^2
  \approx\frac{1}{\sqrt N}\int_{-U}^{U}\re(\mathcal L(u))w_\sigma^U(u)\di u
  =\frac{2\pi}{\sqrt N}\int_\R\re\left(\F^{-1}f_{\sigma,U}(-x)\right)\varrho(x)\di W(x).
\]
For $\sigma^2$ we calculate the finite sample variance $s_{\sigma^2}^2$ of the linearized stochastic errors using the It\^{o} isometry
\begin{align}
  s_{\sigma^2}^2&=\frac{1}{N}\mathbb{E}\left[\left(\int_{-U}^{U} \mathrm{Re}(\mathcal{L}(u))w_\sigma^U(u)\di u\right)^2\right]
  =\frac{4\pi^2}{N}\mathbb{E}\left[\left(\int_\R\re\left(\F^{-1}f_{\sigma,U}(-x)\right)\varrho(x)\di W(x)\right)^2\right]\nonumber\\
  &=\frac{4\pi^2}{N}\int_\R\Big(\re\big(\F^{-1}f_{\sigma,U}(-x)\big)\varrho(x)\Big)^2\di x.\label{sigmaVar}
\end{align}
Similar results for $\hat\gamma_{fa},\hat\lambda$ and $\hat\nu(x_0),x_0\in\R,$ are derived in the appendix. The corresponding finite sample variances $s_{\gamma_{fa}}^2, s_\lambda^2$ and $s_{\nu(x_0)}^2$ are given by \eqref{eqS2gamma}, \eqref{eqS2lambda} and \eqref{eqS2nu}, respectively. In contrast to the central limit theorems of \cite{soehl:2012}, we do not have to distinguish between the cases $x_0=0$ and $x_0\neq0$ in the finite sample analysis. In the \SD model the estimators $\hat\gamma_{sd}$ and $\hat\alpha$ have exactly the same structure such that the above analysis applies in this context, too. Their finite sample variances $s_{\gamma_{sd}}^2$ and $s_{\alpha}^2$ are given by \eqref{eqS2gammaSD} and \eqref{eqS2alpha} in the appendix. Note that the characteristic function $\varphi_T$ has of cause a different shape in the \SD scenario. The pointwise variances $s^2_{k(x_0)}$ for the k-function is based on a linearization of $\tilde\psi'(u)-\psi'(u)$ and given by \eqref{eqS2k}.

To construct confidence intervals for $\theta\in\{\sigma^2, \gamma_{fa}, \lambda, \nu(x_0), \gamma_{sd}, \alpha,k(x_0)\}$, we need an estimate $\hat s_{\theta}$ of the standard deviation. To this end, the function $f_{q,U}$ has to be replaced by its empirical version. Since the only unknown quantity involved is $\varphi_T$, it suffices to plug in an estimator for the characteristic function. Either one uses the L\'evy-Khintchine representation~\eqref{eqLevyKhint} replacing the true L\'evy triplet by their estimators or $\varphi_T$ is estimated by the empirical version of \eqref{eqPsi} and \eqref{eqPsi-i}. We will follow the latter approach because the estimate is independent of the cut--off value $U$ and thus may lead to more stable results. To compute the noise function $\varrho$, the density $h$ of the distribution of the strikes is necessary but not known to the practitioner. It can be estimated from the observation points $(x_j)_{j=1,\dots,N}$ using some standard density estimation method. We will apply a triangular kernel estimator, where the bandwidth is chosen by Silverman's rule of thumb. Due to the asymptotic normality proved by \citet{soehl:2012}, the $(1-t)$--confidence intervals for a level $t>0$ are then given by
\begin{equation}\label{intervals}
  I_{\theta}:=[\hat \theta-\hat s_{\theta} q_{t/2},\hat \theta+\hat s_{\theta} q_{t/2}],
\end{equation}
where $q_t$ denotes the $(1-t)$--quantile of the standard normal distribution. Naturally, both the estimator $\hat\theta$ and the size of the confidence set, determined by $\hat s_\theta$, depend the choice of the cut--off value $U$. In particular, it reflects the bias--variance trade--off of the estimation problem: Small values of $U$ lead to a strong smoothing such that the interval \eqref{intervals} will be small but there might be a significant bias. Using larger $U$, the confidence intervals become wider but the deterministic error reduces. Therefore, only by undersmoothing the interval \eqref{intervals} has asymptotically the level $t$.\par

In practice we are rather interested in the parameter $\sigma$ than in its square. Applying the delta method, the finite sample variance of the estimator $\hat\sigma:=\sqrt{\hat\sigma^2}$ is given by $s^2_\sigma=\frac{1}{4}s^2_{\sigma^2}\sigma^{-2}$ and thus its empirical version is $\hat s_\sigma^2=\frac{1}{4}\hat s^2_{\sigma^2}(\hat\sigma^2)^{-1}$. This allows to construct confidence intervals for $\hat\sigma$, too.

We examine the performance of the confidence intervals by simulations from the Merton model and from the variance gamma model with parameters as in Examples~\ref{exMerton} and \ref{exVG}, respectively. As in Section~\ref{sec:FAsim}, the interest rate is chosen as $r=0.06$ and the time to maturity as $T=0.25$. We simulate $N=100$ strike prices and take the relative noise level to be $\tau=0.01$. To coincide with the theory, we interpolate the corresponding European call prices linearly. In the real data application in Section~\ref{sec:empirical} we will take advantage of the B-spline interpolation.

\begin{table}[tp]\centering
  \begin{tabular}{c|cccc|ccc}\hline
            & \multicolumn{4}{c|}{\FA} & \multicolumn{3}{c}{\SD}\\
            & $\sigma^2$  & $\gamma_{fa}$ & $\lambda$ & $\nu(x_0)$ & $\gamma_{sd}$ & $\alpha$ & $k(x_0)$\\\hline
      $U$       & 54 & 50 & 46 & 26 & 35 & 45 & 3\\
      $t=0.5$   & 53\% & 48\% &  43\% & 48\% & 52\% & 49\% & 51\%\\
      $t=0.05$  & 94\% & 93\% &  81\% & 91\% & 100\% & 99\% & 96\%\\\hline
  \end{tabular}
  \caption{Approximate coverage probabilities of $(1-t)$--confidence intervals from a Monte Carlo simulation with 1000 iterations and fixed cut--off values $U$. The confidence intervals of $\nu$ and $k$ are evaluated at $x_0=-0.2$.}\label{tab:coverage}
\end{table}
We asses the performance of the confidence intervals \eqref{intervals} with levels $t=0.5$ and $t=0.05$ in a Monte Carlo simulation with 1000 iterations in each model. The cut--off values are fixed for any quantity and larger than the oracle choice of $U$. This ensures that the bias is indeed negligible. As a rule of thumb the cut--off values for the confidence sets can be chosen to be 4/3 of the oracle cut--off value. We approximate the coverage probabilities of the confidence sets by the percentage of confidence intervals which contain the true value. Table~\ref{tab:coverage} gives the chosen cut--off values and the approximate coverage probabilities. Further simulations show that for sufficiently small levels, for instance $t=0.05$, the confidence intervals have a reasonable size for a wide range of cut--off values. However, in the \FA setting the parameter $\lambda$ falls a bit out of the general picture and the confidence sets with level $t=0.05$ are slightly to large in the \SD scenario.

\begin{figure}[tp]\centering
    \includegraphics[width=7.2cm]{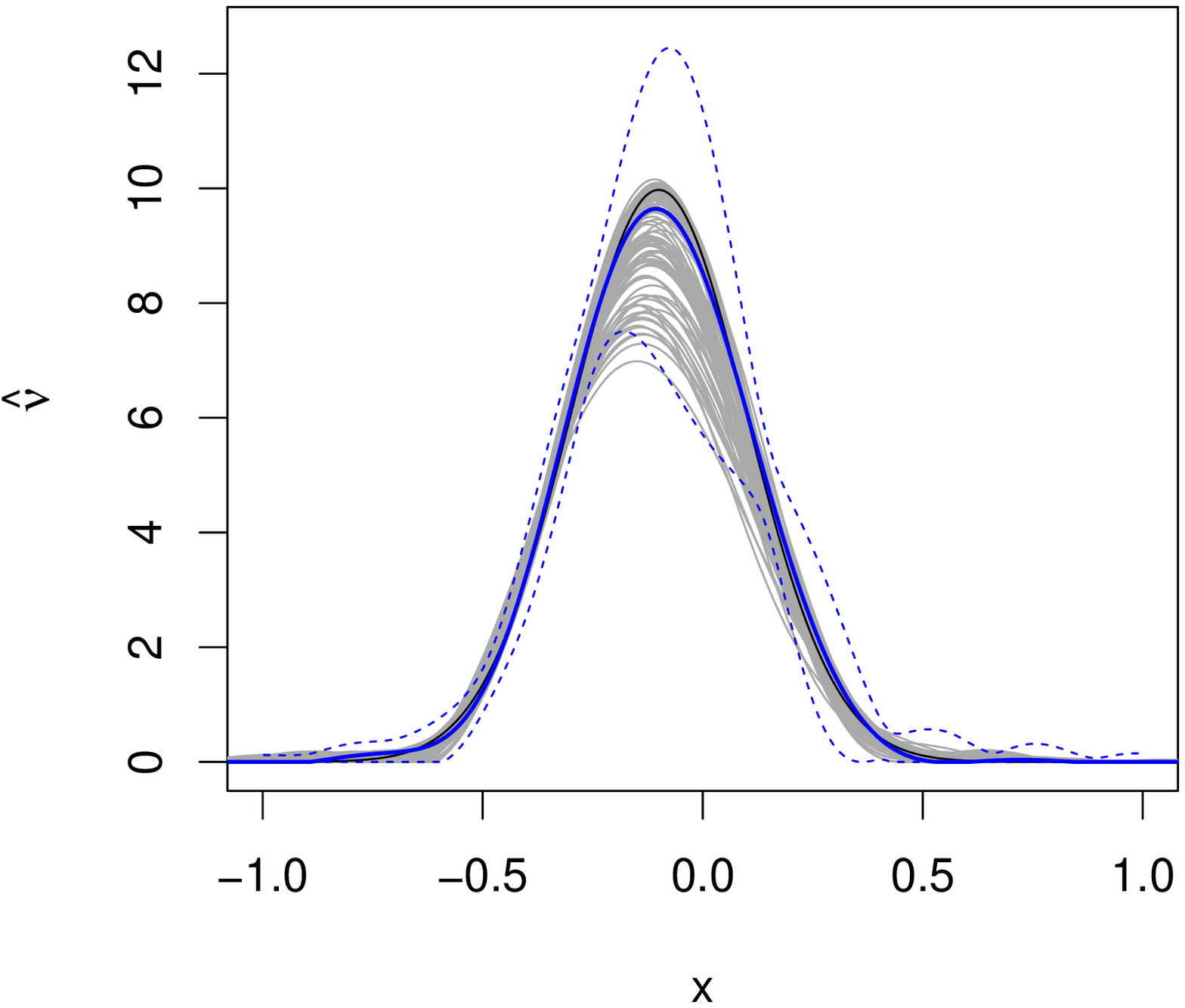}
    \includegraphics[width=7.2cm]{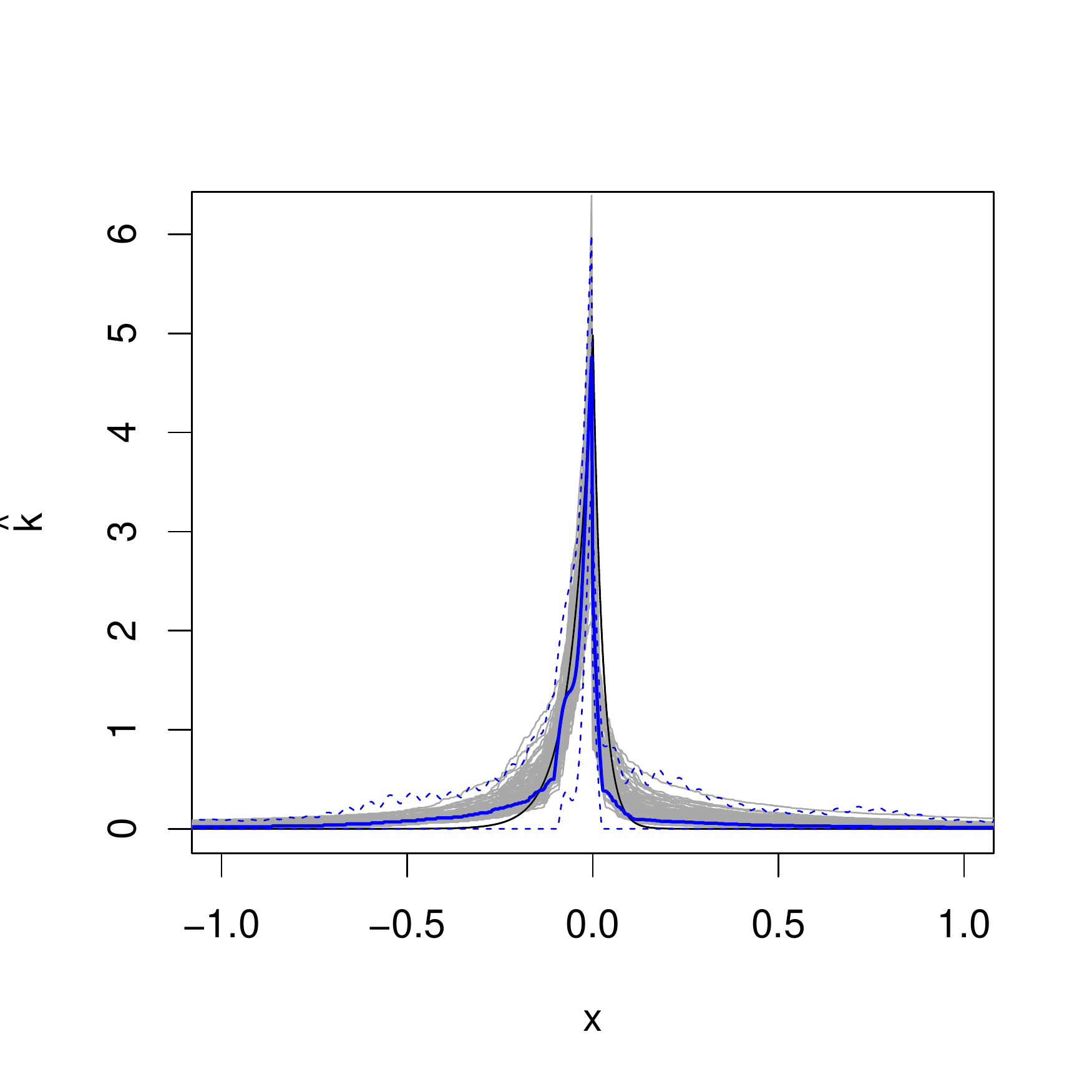}
    \begin{minipage}[t]{7.2cm}
      \caption{True \textit{(black, solid)} and estimated \textit{(blue, bold)} L\'evy density with pointwise 95\% confidence intervals \textit{(blue, dashed)}, using the oracle cut--off value $U=19$. Additional 100 estimated L\'evy densities \textit{(grey)} from a Monte Carlo simulation of the Merton model.}\label{confidenceNu}
    \end{minipage}
    \begin{minipage}{0.2cm}$\,$\end{minipage}
    \begin{minipage}[t]{7.2cm}
      \caption{True \textit{(black, solid)} and estimated \textit{(blue, bold)} k-function with pointwise 95\% confidence intervals \textit{(blue, dashed)}, using the oracle cut--off value $U=2.8$. Additional 100 estimated k-function \textit{(grey)} from a Monte Carlo simulation of the variance gamma model.}\label{confidenceK}
    \end{minipage}
\end{figure}
Based on simulations in the Merton and the variance gamma model, Figures~\ref{confidenceNu} and \ref{confidenceK} illustrate the true L\'evy density and k-function, respectively, their estimators with oracle choice of the cut--off values and the corresponding pointwise 95\% confidence intervals. Almost everywhere the true function is contained in the confidence intervals. Moreover, another 100 estimators from further Monte Carlo iterations are plotted. The graphs show that the confidence intervals describe well the deviation of the estimated jump densities. The negative bias around zero might come from the smoothing which naturally tends to smooth out peaks, cf. \cite[Chap. 5.3]{haerdle:1990}.

\section{Empirical study}\label{sec:empirical}

\begin{figure}[tp]\centering
  \includegraphics[width=9cm]{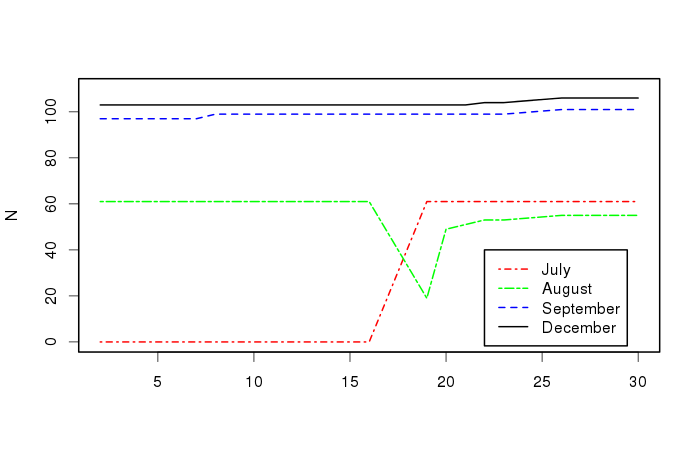}
  \caption{Number of observed prices of put and call options during May, 2008.}\label{figNumber}
\end{figure}

We apply the calibration methods to a data set from the Deutsche B\"orse database Eurex\footnote{provided through the SFB 649 ``Economic Risk''}. It consists of settlement prices of European put and call options on the DAX index from May 2008. Therefore, the prices are observed before the latest financial crises and thus the market activity is stable. The interest rate $r$ is chosen for each maturity separately according to the put--call parity at the respective strike prices. The expiry months of the options are between July and December, 2008, and thus the time to maturity $T$, measured in years, reaches from two to seven months. The number of our observations $N$ is given in Figure~\ref{figNumber} and lays around 50 to 100 different strikes for each maturity and trading day.\par
To apply the confidence intervals \eqref{intervals} of Section~\ref{sec:confidence}, we need the noise function $\varrho$ from \eqref{varrho}. By a rule of thumb we assume $\delta$ to be 1\% of the observed prices $\mathcal O(x_j)$ \cite[cf.][p. 439]{contTankov:2004b}. All other unknown quantities are estimated as discussed above.\par

\subsection{Comparison of \FA and \SD}
\begin{figure}[tp]\centering
 \includegraphics[width=14cm]{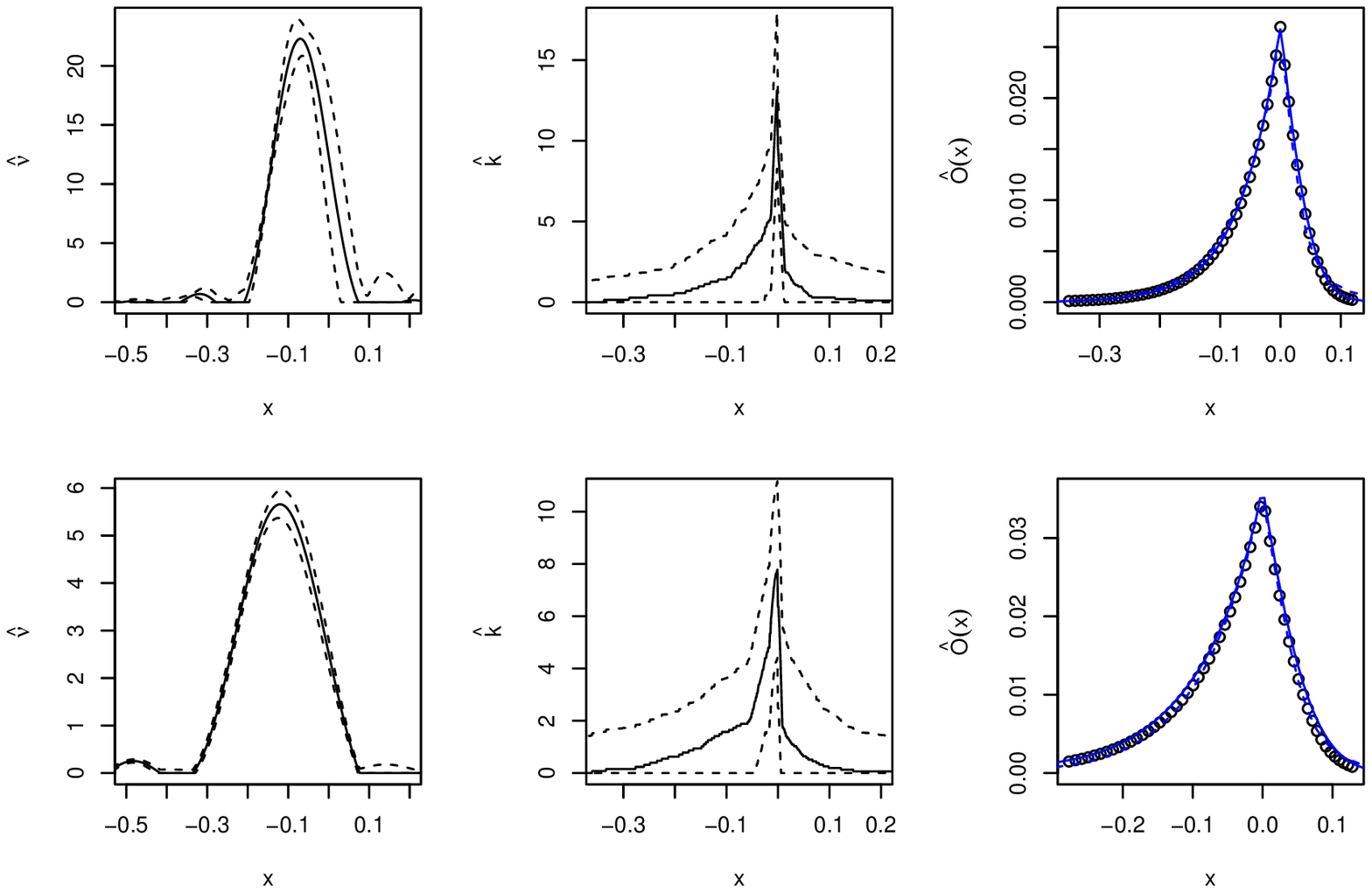}
 \includegraphics[width=14cm]{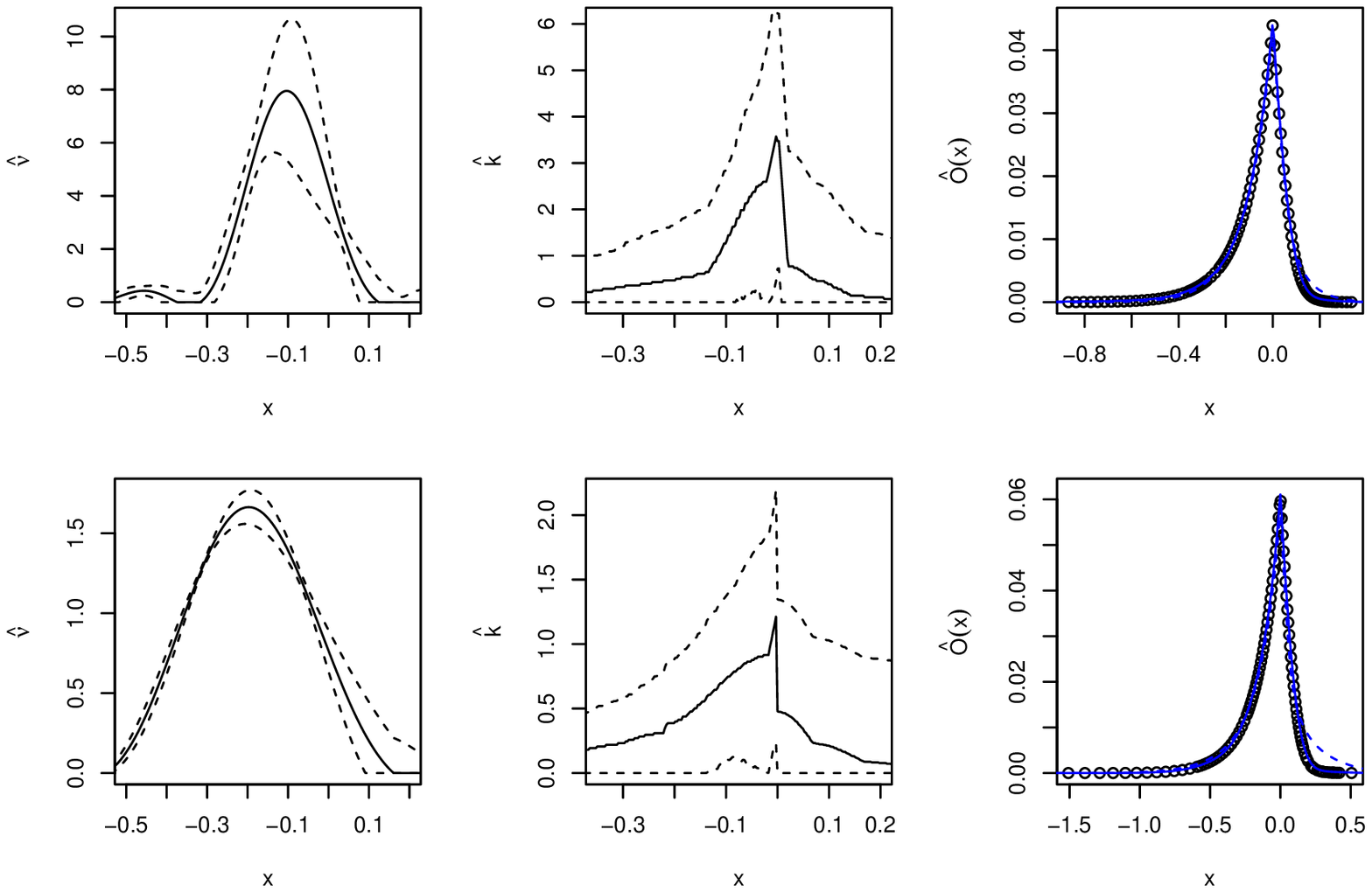}
 \caption{Estimated jump densities \textit{(left)}, k--functions \textit{(center)} with pointwise 95\% confidence intervals as well as calibrated option functions in the \FA \textit{(right, solid)} and \SD \textit{(right, dashed)} setting and given data from May 29, 2008 \textit{(right, points)}. The time to maturity increases from $T=0.136$ \textit{(top)} to $T=564$ \textit{(bottom)}.}\label{fig:May29}
\end{figure}

Let us first focus on one (arbitrarily chosen) day. Hence, we calibrate the option prices of May 29, 2008, with all four different maturities to both, the \FA and the \SD setting. The results are summarized in Table~\ref{tab:May29} and Figure~\ref{fig:May29}. Using the complete estimation of the models, we generate the corresponding option functions $\hat{\mathcal O}$. They are graphically compared to the given data points and we calculate the residual sum of squares $RSS=RSS(U^*)$ as defined in \eqref{eqRSS}. For all maturities both methods yield good fits to the data. However, for longer maturities, especially the calibration of options with seven months to maturity, minor problems occur in the \SD calibration. Although the sample size is larger, the estimated standard deviation is larger for longer maturities in the \SD scenario, too. The calibration at other trading days confirms this weakness of the \SD method for larger $T$. This coincides with the asymptotic analysis of \cite{trabs:2011} where longer durations lead to slower convergence rates of the risk.\par
Moreover, Figure~\ref{fig:May29} shows that the estimated option function $\hat{\mathcal O}$ which results from the \SD calibration does not exactly recover the tails of $\mathcal O$. In all maturities and in both models the L\'evy density has more weight on the negative half line and thus there are more negative jumps than positive ones priced into the options. This coincides with the empirical findings in the literature \citep[see eg,][]{contTankov:2004b}. Due to the positivity correction, the jump densities might look unsmooth where they are close to zero. This problem might be circumvented by adding smoothness constraints. However, the construction of confidence intervals would then be much more difficult. Hence, this topic is left open for further research.\par

In view of the parametric calibration of their CGMY model \citet{Carr:2002} suggested that risk--neutral processes of stocks should be modeled by pure jump processes with finite variation. Now, the nonparametric approach shows that both models the finite activity case and the self--decomposable model are able to reproduce the option data. The finite activity jump-diffusion seem to work even more robust with respect to $T$. Note that in both models the Blumenthal--Getoor index equals to zero which is in contrast to the investigation of high-frequency historical data, where \cite{aitSahaliaJacod:2009} estimated a jump activity index larger than one.\par

\begin{table}[tp]\centering
  \begin{tabular}{ccrrrrrrrr}\hline
    \multirow{3}{*}{} & N & 61 && 55 && 101 && 106&\\
    & T & 0.136 && 0.233 && 0.311 && 0.564 &\\\hline
    \multirow{4}{*}{\FA} & $\hat\sigma$  & 0.110 & (0.0021) & 0.123 & (0.0009) & 0.107 & (0.0030) & 0.124 & (0.0013)\\
    & $\hat\gamma_{fa}$ & 0.221 & (0.0049) & 0.142 & (0.0015) & 0.174 & (0.0050) & 0.105 & (0.0011)\\
    & $\hat\lambda$ & 3.392 & (0.2015) & 1.290 & (0.0176) & 1.823 & (0.1261) & 0.637 & (0.0181)\\
    & $\sqrt{RSS}$ & 0.003 && 0.008 && 0.005 && 0.008&\\\hline
    \multirow{3}{*}{\SD} & $\hat\gamma_{sd}$ & 0.344 & (0.0103) & 0.336 & (0.0136) & 0.302 & (0.3242) & 0.139 & (0.0607)\\
    & $\hat\alpha$ & 8.662 & (0.1534) & 8.677 & (0.2938) & 3.670 & (0.0797) & 5.181 & (1.0030)\\
    & $\sqrt{RSS}$ & 0.007 && 0.006 && 0.011 && 0.029&\\\hline
  \end{tabular}
  \caption{Estimated parameters $\theta$ and estimated standard deviation $\hat s_\theta$ (in brackets) for $\theta\in\{\sigma,\gamma_{fa},\lambda,\gamma_{sd},\alpha\}$ and residual sum of squares using option prices from May 29, 2008, with $N$ observed strikes for each maturity $T$.}\label{tab:May29}
\end{table}

\subsection{\FA across trading days}
The aim of this section is twofold. By considering more than one day we investigate the stability of the \FA estimation procedure. Moreover, calibrating the model across the trading days in May, 2008, shows the development of the model along the time line and with small changes in the maturities. To profit from the higher observation number, we apply the calibration procedure for the \FA case to the options with maturity in September and December.\par
\begin{figure}[tp]\centering
 \includegraphics[width=9cm]{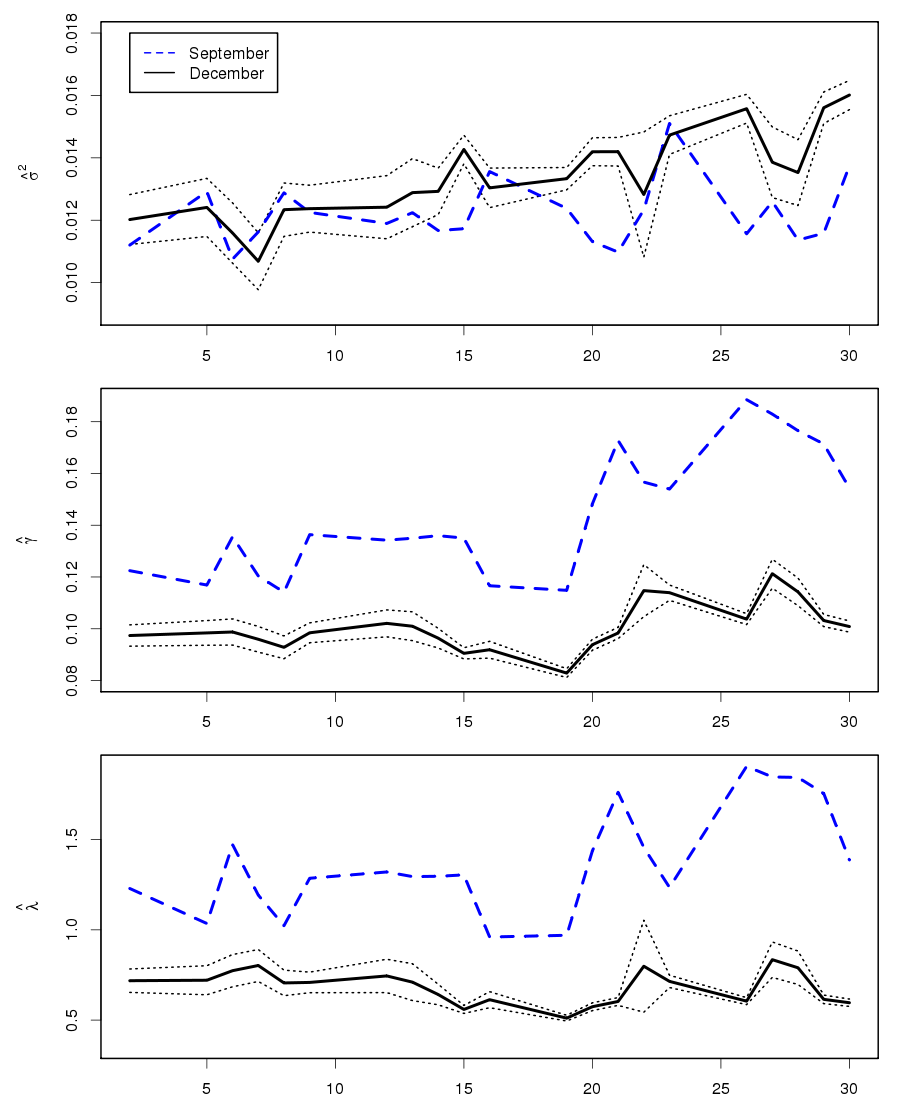}
 \caption{At each market day in May, 2008, estimated $\sigma^2$ \textit{(top)}, $\gamma$ \textit{(center)} and $\lambda$ \textit{(bottom)} from options with maturities in September \textit{(dashed)} and December \textit{(solid)} and confidence intervals \textit{(dotted)} for the latter ones.}\label{fig:paraMay}
\end{figure}

The estimations of the parameters are displayed in Figure~\ref{fig:paraMay}. Note that we do not smooth over time. Furthermore, the 95\% confidence intervals for the December options are shown. The estimated volatility $\hat\sigma$ fluctuates around 0.1 and 0.12. The confidence sets imply that there is no significant difference between the two maturities. Both $\hat{\gamma}_{fa}$ and $\hat{\lambda}$ decrease for higher durations: On the one hand the curves of December lay significantly below the ones of September, on the other hand the graphs have a slight positive trend with respect to the time axis, which means with smaller time to maturity. Keeping in mind that the implied volatility in the Black--Scholes model typically decreases for longer time to maturity, this lower market activity is reproduced by smaller jump activities in our calibration while the volatility is relatively stable.\par
\begin{figure}[tp]\centering
 \includegraphics[width=7cm]{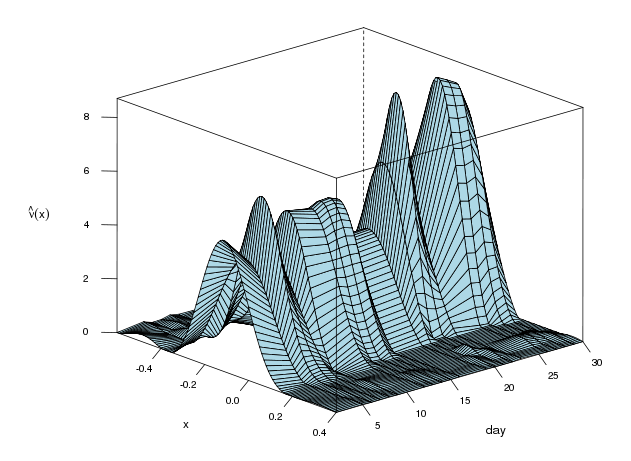}
 \includegraphics[width=7cm]{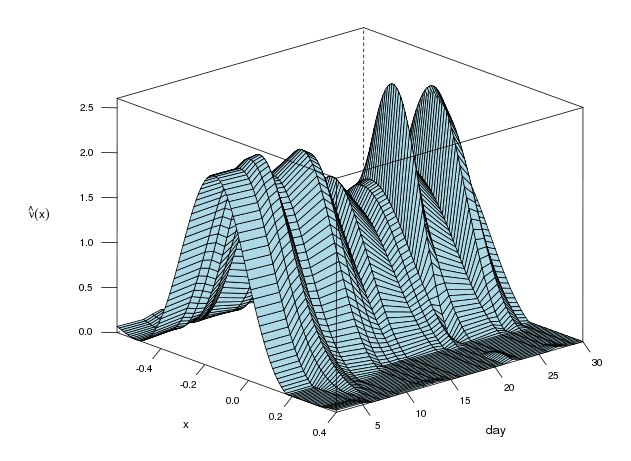}
 \caption{Estimation of $\nu$ for maturity in September \textit{(left)} and December (\textit{right)}.}\label{fig:nuMay}
\end{figure}

Figure~\ref{fig:nuMay} displays the estimated jump densities. All jump measures have a similar shape which is in line with real data calibration of \cite{reiss2:2006}. In contrast to \cite{contTankov:2004} the densities are unimodal or have only minor additional modes in the tails, which may be artefacts of the spectral calibration method. The tails of $\hat\nu$ do not differ significantly, while the different heights reflect the development of the jump activities $\hat\lambda$. There is an obvious trend to small negative jumps in all data sets, which is in line with the stylized facts of option pricing models. The calibration is stable for consecutive market days.\par

\section{Conclusions}\label{sec:conclusion}
To reduce the model misspecification it is reasonable to use a nonparametric model for option pricing. However, the nonlinear inverse problem, which occurs by calibrating the model, is more difficult to solve than parametric calibration problems and needs nonstandard algorithms. We could improve the existing spectral calibration procedures for the finite activity \FA L\'evy model and the self--decomposable \SD L\'evy model. Owing to the fast Fourier transform, the method is computationally fast and admits convincing results in simulations and real data applications. Determining the finite sample variances of the linearized estimators, we obtain confidence sets, which allow a precise analysis of the estimation errors.\par
Our empirical investigations show that both models can be calibrated well to European option prices. However, \FA is more suitable for longer maturities. Using the derived confidence intervals, we can observe significant changes of the \FA model over time. While the volatility has no systematic trend, the jump activities decrease for longer maturities and thus the L\'evy densities become flatter.\par
To avoid misspecification of the model, we are convinced that the nonparametric approach should be pushed forward theoretically and in practice, in particular, in view the high number of available observations in highly liquid markets. Of further interest would be extensions of the method to models whose jump part is not of finite variation as well as the application to hedging and risk management problems.

\appendix
\section{Appendix}
Starting with the finite activity model, the confidence intervals for $\gamma$ and $\lambda$ are based on the finite sample variances of the corresponding linearized stochastic errors. With $f_{\gamma_{fa},U}(u):=w_{\gamma_{fa}}^U(u)iu(1+iu)/(T\varphi_T(u-i))$ and $f_{\lambda,U}(u):=w_\lambda^U(u)iu(1+iu)/(T\varphi_T(u-i))$ we obtain by definitions \eqref{gammaHat} and \eqref{lambdaHat} and the same arguments as in Section~\ref{sec:confidence}
\begin{align*}
  \Delta\hat\gamma_{fa}&:=\hat\gamma_{fa}-\gamma
  \approx-\Delta\hat\sigma^2+\int_\R\im\big(\Delta\tilde\psi_{-i}(u)\big)w_{\gamma_{fa}}^U(u)\di u\\
  &\approx\frac{2\pi}{\sqrt N}\int_\R\Big(-\re\big(\F^{-1}f_{\sigma,U}(-x)\big) +\im\big(\F^{-1}f_{\gamma_{fa},U}(-x)\big)\Big)\varrho(x)\di W(x),\\
  \Delta\hat\lambda&:=\hat\lambda-\lambda
  \approx\tfrac{1}{2}\Delta\hat\sigma^2+\Delta\hat\gamma_{fa}-\int_\R\re\big(\Delta\tilde\psi_{-i}(u)\big)w_\lambda^U(u)\di u\\
  &\approx\frac{2\pi}{\sqrt N}\int_\R\Big(-\re\big(\tfrac{1}{2}\F^{-1}f_{\sigma,U}(-x)+\F^{-1}f_{\lambda,U}(-x)\big)+\im\big(\F^{-1}f_{\gamma_{fa},U}(-x)\big)\Big)\varrho(x)\di W(x).
\end{align*}
Therefore, the finite sample variances are given by
\begin{align}
  s_{\gamma_{fa}}^2&=\frac{4\pi^2}{N}\int_\R\Big(-\re\big(\F^{-1}f_{\sigma,U}(-x)\big) +\im\big(\F^{-1}f_{\gamma_{fa},U}(-x)\big)\Big)^2\varrho^2(x)\di x,\label{eqS2gamma}\\
  s_\lambda^2&=\frac{4\pi^2}{N}\int_\R\Big(-\re\big(\tfrac{1}{2}\F^{-1}f_{\sigma,U}(-x)+\F^{-1}f_{\lambda,U}(-x)\big)
  +\im\big(\F^{-1}f_{\gamma_{fa},U}(-x)\big)\Big)^2\varrho^2(x)\di x.\label{eqS2lambda}
\end{align}
The estimator $\hat\nu(x_0), x_0\in\R,$ in \eqref{nuHat} involves $\tilde\psi$ instead of $\tilde\psi_{-i}$. Hence, the confidence intervals for $\nu(x_0)$ are based on the linearization
\[
  \Delta\tilde\psi:=\tilde\psi-\psi
  \approx-\frac{u(u+i)}{T\varphi_T(u)}(\F\tilde{\mathcal O}-\F\mathcal O)(u+i)
  \approx-\frac{1}{\sqrt N}\frac{u(u+i)}{T\varphi_T(u)}\int_\R e^{iux-x}\varrho(x)\di W(x)
  =:\frac{1}{\sqrt N}\mathcal L_\nu(u).
\]
Defining $f_{\nu,U}(u):=-w_\nu^U(u)u(u+i)/(T\varphi_T(u))$ and writing for brevity $g^{(m)}_U(x_0):=\mathcal F^{-1}[u^mw_\nu^U(u)](x_0)$ with $m\in\{0,1,2\}$, the dominating stochastic error term of $\hat\nu(x_0)$ is then given by \citep[cf.][(6.3)]{soehl:2012}
\begin{align*}
  \Delta\hat\nu(x_0):=&\hat\nu(x_0)-\nu(x_0)\\
  \approx&\frac{1}{\sqrt N}\Big(\frac{1}{2\pi}\int_\R e^{-iux_0}\mathcal L_\nu(u)w_\nu^U(u)\di u
  +\frac{\Delta\hat\sigma^2}{2}g_U^{(2)}(x_0)-i\Delta\hat\gamma_{fa} g_U^{(1)}(x_0)+\Delta\hat\lambda g_U^{(0)}(x_0)\Big)\\
  \approx&\frac{2\pi}{\sqrt N}\int_\R\Big(\frac{e^{-x}}{2\pi}\mathcal F^{-1}f_{\nu,U}(x_0-x)
  +\re\big(\F^{-1}f_{\sigma,U}(-x)\big)\big(\tfrac{1}{2}g_U^{(2)}(x_0)+ig_U^{(1)}(x_0)-\tfrac{1}{2}g_U^{(0)}(x_0)\big)\\
  &+\im\big(\F^{-1}f_{\gamma_{fa},U}(-x)\big)\big(-ig_U^{(1)}(x_0)+g_U^{(0)}(x_0)\big)
  -\re\big(\F^{-1}f_{\lambda,U}(-x)\big)g^{(0)}_U(x_0)\Big)\varrho(x)\di W(x),
\end{align*}
where we note that $g_U^{(0)}, g_U^{(2)}$ are purely real and $g_U^{(1)}$ has only an imaginary part by the symmetry of $w_\nu^U$. Hence, the variance of the linearized stochastic error of $\hat\nu(x_0)$ is given by
\begin{align}
  s_{\nu(x_0)}^2=&\frac{4\pi^2}{N}\int_\R\Big(\frac{e^{-x}}{2\pi}\mathcal F^{-1}f_{\nu,U}(x_0-x)
  +\re\big(\F^{-1}f_{\sigma,U}(-x)\big)\big(\tfrac{1}{2}g_U^{(2)}(x_0)+ig_U^{(1)}(x_0)-\tfrac{1}{2}g_U^{(0)}(x_0)\big)\notag\\
  &+\im\big(\F^{-1}f_{\gamma_{fa},U}(-x)\big)\big(-ig_U^{(1)}(x_0)+g_U^{(0)}(x_0)\big)
  -\re\big(\F^{-1}f_{\lambda,U}(-x)\big)g^{(0)}_U(x_0)\Big)^2\varrho^2(x)\di x\label{eqS2nu}.
\end{align}

Let us now consider the self--decomposable model. Compared to the analysis of $\hat\sigma^2$ in Section~\ref{sec:confidence}, the stochastic errors of the estimators $\hat\gamma_{sd}$ and $\hat\alpha$ only differ in the weight functions and the underlying form of the characteristic function $\varphi_T$. We obtain
\begin{align*}
  \Delta\hat\gamma_{sd}&:=\hat\gamma_{sd}-\gamma
  \approx\int_\R\im\big(\Delta\tilde\psi_{-i}(u)\big)w_{\gamma_{sd}}^U(u)\di u
  \approx\frac{2\pi}{\sqrt N}\int_\R\im\Big(\F^{-1}f_{\gamma_{sd},U}(-x)\Big)\varrho(x)\di W(x),\\
  \Delta\hat\alpha&:=\hat\alpha-\alpha
  \approx\int_\R\re\big(\Delta\tilde\psi_{-i}(u)\big)w_\alpha^U(u)\di u
  \approx\frac{2\pi}{\sqrt N}\int_\R\re\Big(\F^{-1}f_{\alpha,U}(-x)\Big)\varrho(x)\di W(x)
\end{align*}
with $f_{\gamma_{sd},U}(u):=w_{\gamma_{sd}}^U(u)iu(1+iu)/(T\varphi_T(u-i))$ and $f_{\alpha,U}(u):=w_\alpha^U(u)iu(1+iu)/(T\varphi_T(u-i))$. The finite sample variances are thus given by
\begin{align}
  s_{\gamma_{sd}}^2&=\frac{4\pi^2}{N}\int_\R\Big(\im\big(\F^{-1}f_{\gamma_{sd},U}(-x)\big)\Big)^2\varrho^2(x)\di x\quad\text{and}\label{eqS2gammaSD}\\
  s_\alpha^2&=\frac{4\pi^2}{N}\int_\R\Big(\re\big(\F^{-1}f_{\alpha,U}(-x)\big)\Big)^2\varrho^2(x)\di x.\label{eqS2alpha}
\end{align}
The estimator $\hat k$ is based on $\tilde\psi'$, which is given by~\eqref{eqPsiPrim} with the empirical versions $\F\tilde{\mathcal{O}}$ and $\F[x\tilde{\mathcal{O}}]$. 
We define $\F[x\tilde{\mathcal{O}}](u):=\F[x\di Z](u)=\F[x\mathcal{O}](u)+N^{-1/2}\int_{\R}xe^{iux}\varrho(x)\di W(x)$.
In view of \citep[p.~21]{trabs:2011} and~\eqref{eqPsiPrim} the linearized stochastic error is given by
\begin{align*}
  &\Delta\tilde\psi'(u):=\tilde\psi'(u)-\psi'(u)\\
  \approx&\frac{1}{T\varphi_T(u)}\big((u-iu^2)(\F[x\tilde{\mathcal O}]-\F[x\mathcal O])(u+i)-(2u+i)(\F\tilde{\mathcal O}-\F\mathcal O)(u+i)\big)\\
  &\hspace{1.5cm}+\frac{\varphi_T'(u)}{T\varphi_T^2(u)}\big(u(u+i)(\F\tilde{\mathcal O}-\F\mathcal O)(u+i)\big)\\
  \approx&\frac{(u-iu^2)}{\sqrt NT\varphi_T(u)}\int_Rxe^{iux-x}\varrho(x)\di W(x)+\Big(\frac{\varphi_T'(u)(u^2+iu)}{\sqrt NT\varphi_T^2(u)}-\frac{2u+i}{\sqrt NT\varphi_T(u)}\Big)\int_Re^{iux-x}\varrho(x)\di W(x)\\
  =:&\frac{1}{\sqrt N}\mathcal L_k(u)
\end{align*}
We define $f_{k,U}^{(1)}(u):=\F W_k(u/U)(u-iu^2)/(T\varphi_T(u))$ as well as $f_{k,U}^{(2)}(u):=\F W_k(u/U)\Big((u^2+iu)\varphi_T'(u)/(T\varphi_T^2(u))-(2u+i)/(T\varphi_T(u))\Big)$ and thus for $x_0>0$
\begin{align*}
  \Delta\hat k(x_0)&:=\hat k(x_0)-k(x_0)
  \approx\frac{1}{\sqrt N}\Big(\frac{-i}{2\pi}\int_\R e^{-iux_0}\mathcal L_k(u)\F W_k(u/U)\di u-\Delta\hat\gamma_{sd}UW_k(Ux_0)\Big)\\
  &\approx\frac{1}{\sqrt N}\int_\R\Big(-ixe^{-x}\F^{-1}f_{k,U}^{(1)}(x_0-x)-ie^{-x}\F^{-1}f_{k,U}^{(2)}(x_0-x)\Big)\varrho(x)\di W(x).
\end{align*}
Note that $W_k(Ux_0)=0$ for $x_0>0$ because $\supp W_k\subset(-\infty,0]$.
Consequently,
\begin{align}
  s_{k(x_0)}^2=&\frac{1}{N}\int_\R\Big(ixe^{-x}\F^{-1}f_{k,U}^{(1)}(x_0-x)+ie^{-x}\F^{-1}f_{k,U}^{(2)}(x_0-x)\Big)^2\varrho^2(x)\di x\label{eqS2k}
\end{align}
and similarly for negative $x_0$.

\bibliography{reference}
\bibliographystyle{chicago}

\end{document}